\pdfoutput=1

\documentclass[aps,prd,tightenlines,floatfix,twocolumn]{revtex4}
\usepackage{graphicx}
\usepackage{wrapfig}
\usepackage[all]{xy}
\newcommand{\nc}{\newcommand}

\nc{\beq}{\begin{equation}}
\nc{\eeq}{\end{equation}}
\nc{\bea}{\begin{eqnarray}}
\nc{\eea}{\end{eqnarray}}
\nc{\n}{\nonumber \\}

\usepackage[usenames]{color}
\definecolor{DarkGreen}{rgb}{0.0,0.5,0.0}

\begin{document}  

\date{October 19, 2011} 
\title{Probing dark matter streams with CoGeNT}
\author{Aravind Natarajan}
\email{anat@andrew.cmu.edu}
\affiliation{McWilliams Center for Cosmology, Carnegie Mellon University, Department of Physics, 5000 Forbes Ave., Pittsburgh PA 15213, USA}

\author{Christopher Savage}
\email{savage@fysik.su.se}
\affiliation{ The Oskar Klein Centre for Cosmoparticle Physics, Department of Physics, Stockholm University, AlbaNova, SE-10691 Stockholm, Sweden }

\author{Katherine Freese}
\email{ktfreese@umich.edu}
\affiliation{ Michigan Center for Theoretical Physics, University of Michigan, Ann Arbor, MI 48109, USA }

\begin{abstract}
We examine the future sensitivity of CoGeNT to the presence of dark matter streams and find that
consideration of streams in the data may lead to differences in the interpretation of the
results. We show the allowed particle mass and cross section for different halo parameters, assuming spin-independent elastic scattering. As an example, we choose a stream with the same velocity profile as that of the
Sagittarius stream (and in the Solar neighborhood) and find that,
 with an exposure of $\sim$ 10 kg year,
the CoGeNT results can be expected to
exclude the SHM-only halo in favor of an SHM+stream halo at the 95\% (99.7\%)
confidence level 
provided the stream contributes 3\% (5\%) of the local dark matter density.  The presence of a significant stream component may result in incorrect estimates of the particle mass and cross section unless the presence of the stream is taken into account. We conclude that the CoGeNT experiment is sensitive to streams and care should be taken to include the possibility of streams when analyzing experimental results.
\end{abstract}

\maketitle

\section{Introduction}

The nature of the dark matter (DM) that comprises the bulk of the mass in the Universe is
one of the longest outstanding problems in all of physics.   Weakly Interacting Massive Particles (WIMPs) with weak scale cross sections and masses in the GeV to TeV range are among the best motivated dark matter candidates.
Recent experiments
indicate hints that the DM particle may have been detected.
However, interpretation of the data depends on our understanding of the velocities
of these particles as they pass through the detectors.  The canonical distribution
of WIMPs in the Halo of our Galaxy is the isothermal Maxwell-Boltzmann distribution.  However, the real Galaxy is not likely to be quite so simple.  Numerical simulations suggest that dark matter halos may be triaxial and anisotropic, which can lead to significant differences in the scattering rates and the amplitude and phase of the annual modulation \cite{green1, green2}.


Direct detection experiments are sensitive to the form of
the local dark matter velocity distribution \cite{str1, str2, str3, str4}, and the
results may be significantly modified if a cold stream of dark matter
particles exists in the solar neighborhood. Tidal disruption of dwarf
satellites is expected to produce such cold streams. The late infall
of dark matter onto the Galaxy is also expected to result in cold
streams contributing $\sim$ a few percent to the local dark matter
density  \cite{sikivie1, sikivie2, sikivie3}.
Cosmological N-body simulations from the Aquarius project \cite{white1, white2} 
show the presence of tidal streams in the solar neighborhood at the level of $\sim 1\%$ of the local dark matter density, at $\sim 20\%$ probability. Larger stream contributions are likely if dense tidal streams pass near the sun's location.  Streams will tend to dominate in the outer halo, rather than the inner halo - as has been found in stellar surveys thus far (for a discussion, see \cite{helmi}).

Recently, \cite{green_new} have studied the effect of major mergers and conclude that massive ($>10^{10} M_\odot$) mergers can lead to observable structure in the velocity distribution of dark matter particles. Authors \cite{sims1, sims2} find that such a merger event is very likely. Previously \cite{sag1, sag2, sag3} showed that in principle such streams should be visible in data from DM detectors by giving rise to a step in the energy spectrum of the count rate:
the count rate would be higher up to a critical energy above which the stream could no
longer contribute to the data.  Future experiments will be sensitive to a wide range of stream velocities.
In particular, as we will show,
the CoGeNT experiment may have, over the next
few years, the sensitivity to detect such streams.

In 2010, the CoGeNT collaboration \cite{cogent_old} reported an
excess of events at low energies, which was interpreted as possibly
due to scattering of dark matter particles with the target
\cite{cogent_old, hooper1,kelso_hooper,chang}.  More recently, CoGeNT
reported \cite{cogent_new} the
detection of an annual modulation \cite{kt1, kt2} in the event rate at the
2.8$\sigma$ level, with a modulation amplitude of $16.6 \pm 3.8$\%. The CRESST-II  collaboration has also seen unexplained events that are compatible with a possible explanation in terms of WIMPs \cite{cresst}.   It is tantalizing to compare these new results with the decade-old annual
modulation seen in the DAMA data \cite{dama1, dama2} which has reached
 $9 \sigma$ confidence level.  All these results indicate a possible $\sim 10$ GeV WIMP mass, 
 yet the issue of compatibility between them is unclear.  Some authors
 claim consistency between the different experimental results
 \cite{hooper_kelso, hooper1, belli, frandsen, fornengo} while others do not \cite{new_cogent_paper1, new_cogent_paper2, new_cogent_paper3, new_cogent_paper4}. Authors \cite{gg} showed that the presence of streams could help to
 reconcile the DAMA results with other experiments.

   Yet more perplexing is the apparent discrepancy
 of the null results of XENON \cite{xenon1, xenon2} and CDMS \cite{cdms, cdms2b} with both the DAMA
 and CoGeNT experiments; we do not address these null results in this paper, and restrict
 our discussion only to DAMA and CoGeNT.
 We further address the issue of compatibility between DAMA and CoGeNT in this paper by extending the discussion, within
 the context of the standard Maxwellian halo, to allow
 variation in the two relevant velocities characterizing the distribution: the escape
 velocity and the dispersion.

The focus of the paper, however, is the search for streams in the CoGeNT data.
The low energy threshold (0.47 keVee) and
excellent energy resolution (0.05 keVee) obtained by CoGeNT are key to
detecting dark matter streams for small particle masses. The planned
upgrade to the CoGeNT experiment (C-4) will consist of 4 detectors, of
approximately 1.3 kg each \cite{juan}, and is expected to start taking
data later this year.  We test the sensitivity of the CoGeNT upgrade
to dark matter streams by performing a number of Monte Carlo
simulations and fitting the results.  We select a dark matter mass
and scattering cross-section consistent with the current CoGeNT results,
assuming that the excess events currently seen by CoGeNT at low
energies is entirely due to scattering of dark matter particles with the
target. We caution the reader that if a significant (and currently
unknown) exponential background exists at low energies, our
conclusions may be altered. We also do not consider possible ways in
which the CoGeNT and DAMA results could be made compatible with the
null results of XENON \cite{xenon1, xenon2} and CDMS \cite{cdms, cdms2b}.

Consider dark matter particles of mass $m_\chi$ scattering elastically
off a nucleus of mass $m_{\rm N}$. The number of recoil events per
unit time, per unit detector mass, and per unit energy, is given by
the formula:
\beq
\frac{dR}{dQ}(t,Q) = \frac{\rho_\chi \sigma_{\rm p} A^2 }{2 m_\chi m^2_{\rm R,p}} F^2(Q) \; \int_{v_{\rm min}(Q)}^\infty dv \,  \frac{f(v)}{v}.
\label{rec}
\eeq
$\rho_\chi$ is the dark matter density at the earth's location,
$\sigma_{\rm p}$ is the spin-independent elastic cross section for
WIMP-proton scattering, $A$ is the atomic mass number and $m_{\rm R,p}
= m_\chi m_{\rm p} / (m_\chi + m_{\rm p})$ is the WIMP-proton reduced
mass. We have assumed here that the WIMP coupling to the proton is the
same as the coupling to the neutron. $F(Q)$ is the form
factor, containing the momentum dependence of the cross section, and
takes the form described in \cite{form_factor1, form_factor2, form_factor3}. $f(v)$ is the
one-dimensional speed distribution of dark matter particles relative
to the detector. It is this term that is sensitive to different halo
models. $v_{\rm min}(Q) = \sqrt{Q m_{\rm N}/2 m^2_{\rm R}}$, is the
minimum velocity a particle must have in order to effect a recoil at
energy $Q$, where $m_{\rm R} = m_\chi m_{\rm N} / (m_\chi + m_{\rm
  N})$ is the WIMP-nucleus reduced mass.

The standard halo model (SHM henceforth) is characterized by a
Maxwell-Boltzmann distribution of velocities in the rest frame of the
Galaxy given by
\bea
f_{\rm SHM}(\vec v_{\rm wh}) &=& \frac{1}{N_{\rm esc}} \,  \frac{\exp \left[ - (\vec v_{\rm wh} / v_0)^2 \right ]}{\pi^{3/2} \; v^3_0 } \;\;\;\;\;  \textrm{for} \;\; |\vec{v}_{\rm wh}| < v_{\rm esc} \n
&=& 0 \;\;\;\;\;  \textrm{otherwise}
\label{max}
\eea
where $v_0$ characterizes the velocity dispersion,
$\vec{v}_{\rm wh}$ is the WIMP velocity relative to the halo,
$v_{\rm esc}$ is the escape velocity, and $N_{\rm esc}$ is a
normalization constant chosen such that $\int dv \, f(v) = 1$.
\begin{figure}[!h]
\begin{center}
\scalebox{0.43}{\includegraphics{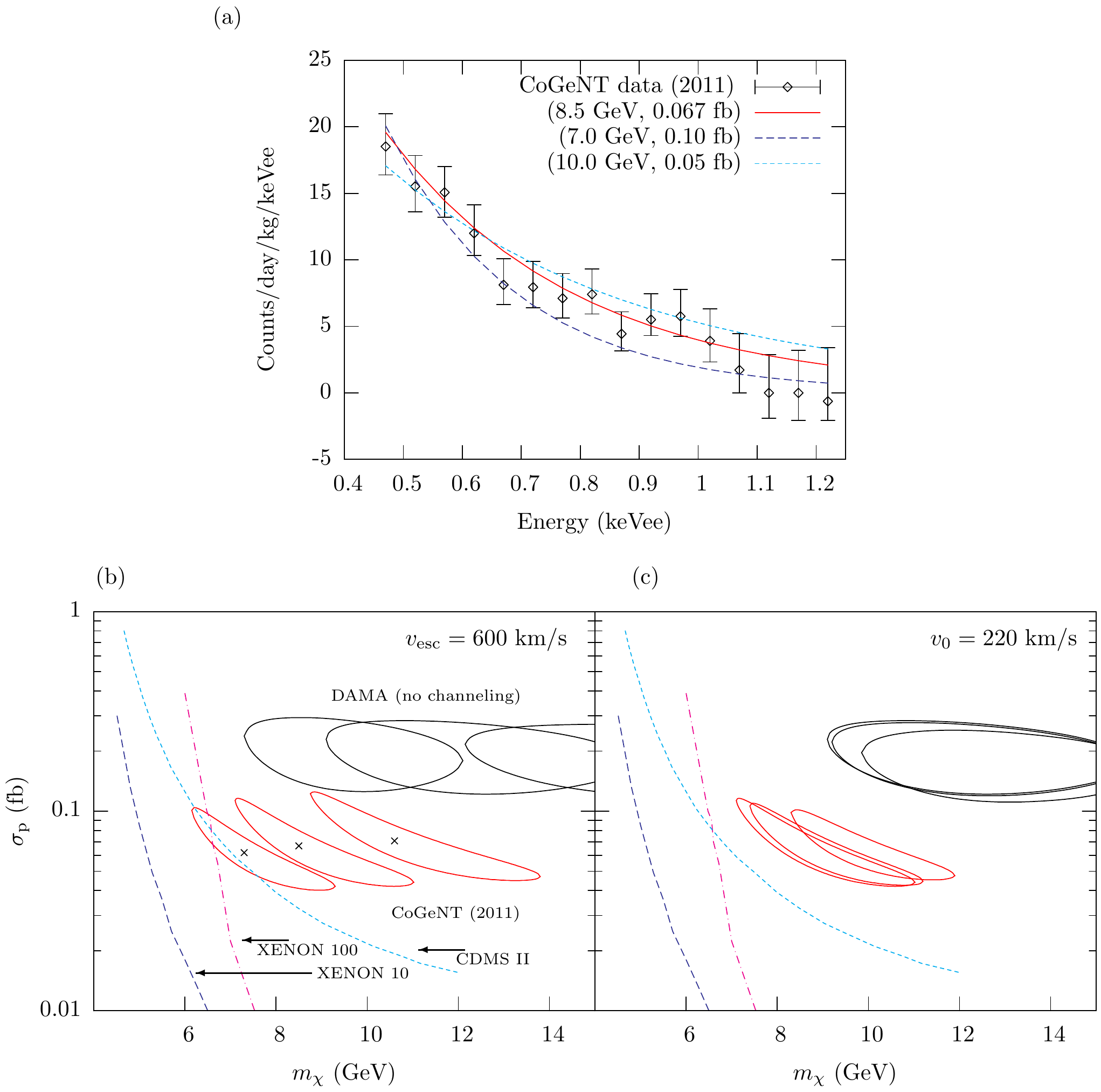}}
\end{center}
\caption{
  (a) The recoil spectrum observed by CoGeNT \cite{cogent_new}
  and several predicted spectra compatible with the observations.
  The solid (red) curve represents the best fit mass and cross section
  for $v_0$=220 km/s, $v_{\rm esc}$=600 km/s.
  The other panels show 3$\sigma$ contours in the
  $(m_\chi, \sigma_{\rm p})$ plane for CoGeNT (solid red) and DAMA
  (solid black) for (b) varying $v_0$ and (c) varying $v_{\rm esc}$
  (see text).  The cross marks indicate the best fit points.  Also
  shown are the exclusion limits from XENON10 (dashed blue), 
  XENON100 (dashed pink), and from the CDMS II low energy analysis (dashed cyan). The XENON10 limit assumes $v_0$ = 230 km/s, $v_{\rm esc}$ = 600 km/s, while the XENON100 limit assumes $v_0$ = 220 km/s, $v_{\rm esc}$ = 544$^{+64}_{-46}$ km/s \cite{xenon1, xenon2}. The CDMS II limit assumes $v_0$ = 220 km/s, $v_{\rm esc}$ = 544 km/s \cite{cdms, cdms2b}.
  \label{fig1} }
\end{figure}

Fig. \ref{fig1}(a) shows the time averaged recoil rate for the low
energy bins, from
\cite{cogent_new}. We have also plotted predicted recoil spectra for
WIMP masses $m_\chi$ = 7, 8.5, and 10 GeV, for the SHM assuming $v_0 =
220$ km/s and $v_{\rm esc} = 600$ km/s. The energy dependence of the
Germanium quenching factor is obtained from the measurements reported
in \cite{texono}. The solid (red) curve represents the best fit
$(m_\chi, \sigma_{\rm p})$, with a $\chi^2$ = 6.3/14 d.o.f. Panel~(b) shows
the 3$\sigma$ allowed contours (solid, red) for $v_0$ = 180, 220, and
260 km/s, from left to right respectively, for an assumed $v_{\rm
  esc}$ = 600 km/s. The cross marks indicate the parameter values that
minimize the value of $\chi^2$ over the 16 lowest energy bins. Also
shown are the 3$\sigma$ contours for the DAMA results
\cite{dama1, dama2,savage2}, ignoring the possibility of channeling, and
assuming a constant quenching factor for Sodium = 0.3 (the recoils off
of Iodine are not significant at these masses). We do not consider varying the Sodium quenching factor here, and caution the reader that the contours will be altered if there is a significant uncertainty in the Sodium quenching factor.  The 90\% exclusion
limits \cite{xenon1, xenon2} for XENON 100 and XENON 10 are plotted. Panel~(c) shows
the variation with $v_{\rm esc}$, for fixed $v_0$ = 220 km/s. Shown
from left to right are 3$\sigma$ contours for $v_{\rm esc}$ = 600,
500, and 400 km/s. Note that while comparing the CoGeNT contours with the XENON exclusion limits, care should be taken to match the halo parameters. We provide CoGeNT contours for different values of $v_0$ and $v_{\rm esc}$, but the XENON and CDMS bounds are for specific values of $v_0$ and $v_{\rm esc}$. Thus the CoGeNT contour for $v_0 = 260$ km/s is likely excluded when compared to the XENON and CDMS exclusion curves for $v_0$ = 260 km/s (not shown). The dark matter density at the sun's location was set to 0.3 GeV/cm$^3$.

\section{Sensitivity to streams}

\begin{figure}[!h]
\begin{center}
\scalebox{0.6}{\includegraphics{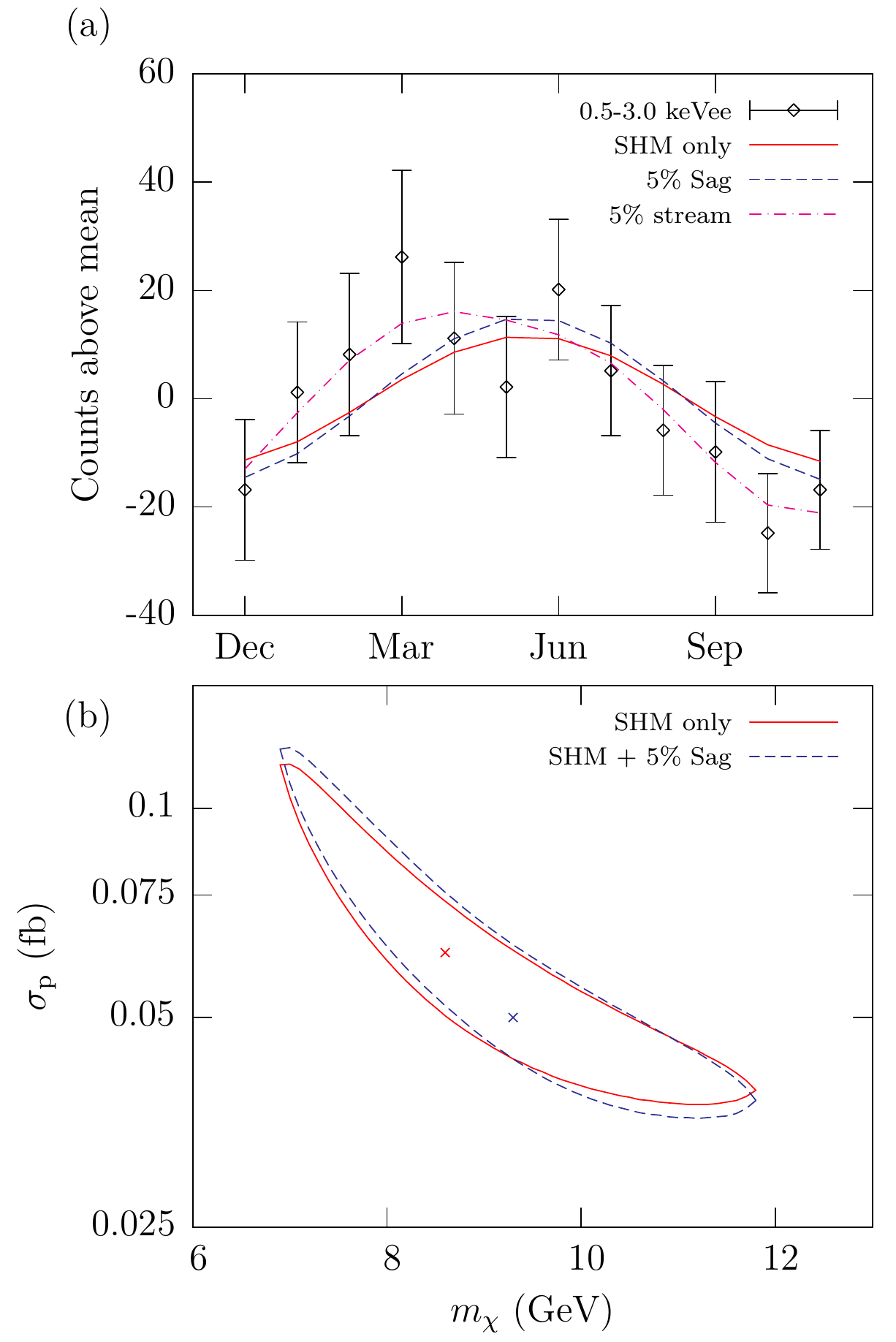}}
\end{center}
\caption{
CoGeNT modulation. Shown in (a) are the first 12 time bins from \cite{cogent_new}, with the mean subtracted. The solid (red) curve is for the SHM. The dashed (blue) curve includes a 5\% contribution from the Sagittarius stream, while the dot-dashed (pink) curve includes a 5\% contribution from an unknown stream. (b) shows the 3$\sigma$ contours combining modulation information with the time averaged recoil rate, with and without a stream contribution, assuming  $v_0$ = 220 km/s and $v_{\rm esc}$ = 600 km/s. The cross marks indicate the best fit parameters.  \label{fig2} }
\end{figure}

We now consider the possibility that in addition to a thermal component of dark matter, there exist dark matter streams,
i.e.\ particles with small or negligible velocity dispersion. As
mentioned in the Introduction, such streams are expected to occur due
to the tidal break-up of small halos, the Sagittarius stream being
a well known example.  The CoGeNT experiment has also measured an annual modulation in the recoil rate, at the $\sim 2.8 \sigma$ level. Fig. \ref{fig2}(a) shows the modulation (mean subtracted) with the expectation for theoretical models with and without streams. Fitting the amplitude to the SHM, we obtain a $\chi^2_{\rm min}$ = 7.8/10 d.o.f, for $m_\chi = 10$ GeV, $\sigma_{\rm p}$ = 0.11 fb. The dashed (blue) curve includes a 5\% contribution from the Sagittarius stream for $m_\chi = 9.3$ GeV, $\sigma_{\rm p}$ = 0.16 fb. With the Sagittarius stream included, the $\chi^2_{\rm min}$ improves to 6.8/10 d.o.f. The pink dot-dashed curve is plotted as an example of an unknown stream that fits the phase of the CoGeNT modulation. A 5\% contribution due to this stream improves the fit significantly, resulting in a $\chi^2_{\rm min}$ = 2.7/7 d.o.f., for a small WIMP mass $m_\chi = 6$ GeV, and a very large cross section $\sigma_{\rm p}$ = 0.69 fb (for the unknown stream, we allow the 3 velocity components to vary, in addition to the mass and cross section). This stream has a velocity relative to the sun $\vec v_{s\odot}$ = (475 km/s, $\theta = 120^\circ, \phi=160^\circ)$, and the co-ordinate system is as defined in \cite{sag2}. For comparison, the Sagittarius stream's velocity relative to the sun \cite{sag1} for $v_0$ = 220 km/s, is (340 km/s, $\theta = 151^\circ, \phi=266^\circ)$. We provide this example merely to illustrate that the presence of streams can have a significant effect on the phase of the annual modulation. More accurate results require better data for the annual modulation.

Fig. \ref{fig2}(b) shows the 3$\sigma$ contours taking into account both the amplitude over one year, and the average recoil rate. The contour for the SHM is very similar to the contour obtained using the time averaged data alone (Fig. \ref{fig1}) owing to the large error bars in the modulation data. The inclusion of a 5\% contribution due to the Sagittarius stream moves the contour only slightly. The best fit values are $m_\chi$ = 8.6 GeV, $\sigma_{\rm p}$ = 0.06 fb, with a $\chi^2_{\rm min}$ = 14.7 with 22 degrees of freedom (12 time bins averaged over energy, 12 energy bins averaged over time, and 2 fitting parameters). With the Sagittarius stream included (5\% contribution), the best fit values are $m_\chi$ = 9.3 GeV, $\sigma_{\rm p}$ = 0.05 fb, with a $\chi^2_{\rm min}$ = 14.9. 

With the inclusion of a stream, the velocity distribution $f(v)$ is modified as: 
\beq
f(v) = \xi \, f_{\rm str} (v) + (1-\xi) f_{\rm SHM}(v)
\label{str}
\eeq
where $\xi$ is the fraction of the dark matter density contributed by
the stream to the local dark matter density. For cold streams, $f_{\rm
  str}(v) \approx \delta(v-v_{\rm str})$ where $v_{\rm str}$ is the
stream speed. A finite velocity dispersion may be accounted for by
replacing the delta function by a Gaussian. For definiteness, let us
consider the Sagittarius stream. The Sagittarius dwarf galaxy is being
tidally disrupted by the Milky Way, resulting in tidal tails.
Previously, the possibility existed that the leading tidal tail passed
through the local neighborhood, allowing for detection by direct
detection experiments (for details, we refer the reader to \cite{sag1}
and references therein).  It is no longer considered likely that the
Sagittarius stream passes near the Earth \cite{Newberg:2011pc}. However, we will use the Sagittarius stream as a case study, to be representative of detecting a dark matter stream of known direction but unknown density. Recently, it was pointed out \cite{purcell_sag} that the infall of the Sagittarius dwarf galaxy may contribute to the formation of the spiral arms of the Milky Way.

We study the effect of adding the Sagittarius stream by performing
Monte Carlo (MC) simulations of future CoGeNT results, taking the stream
to compose a fraction $\xi$ of the local density (which is fixed at
0.3~GeV/cm$^3$).  We only consider the time averaged information, as the present data does not constrain the annual modulation effectively. We consider an exposure of
10~kg-year, which may be obtained in $\sim$ 3 years with
the CoGeNT C-4 detector upgrade.  We assume a known
background that is modeled by a constant plus a double Gaussian, as
done in previous works \cite{chang}. The constant and the heights of
the Gaussian peaks are obtained from \cite{cogent_new}. We include the
known background in all our MC simulations.
For the halo, we take $v_{\rm 0}$ = 220~km/s and $v_{\rm esc}$ = 600~km/s
and we take for our fiducial dark matter mass and cross-section
$m_\chi$ = 10~GeV and $\sigma_{\rm p}$ = 0.05~fb.
These values are consistent with a dark matter signal interpretation of
the excess CoGeNT events.

\begin{figure}[!h]
\begin{center}
\scalebox{0.45}{\includegraphics{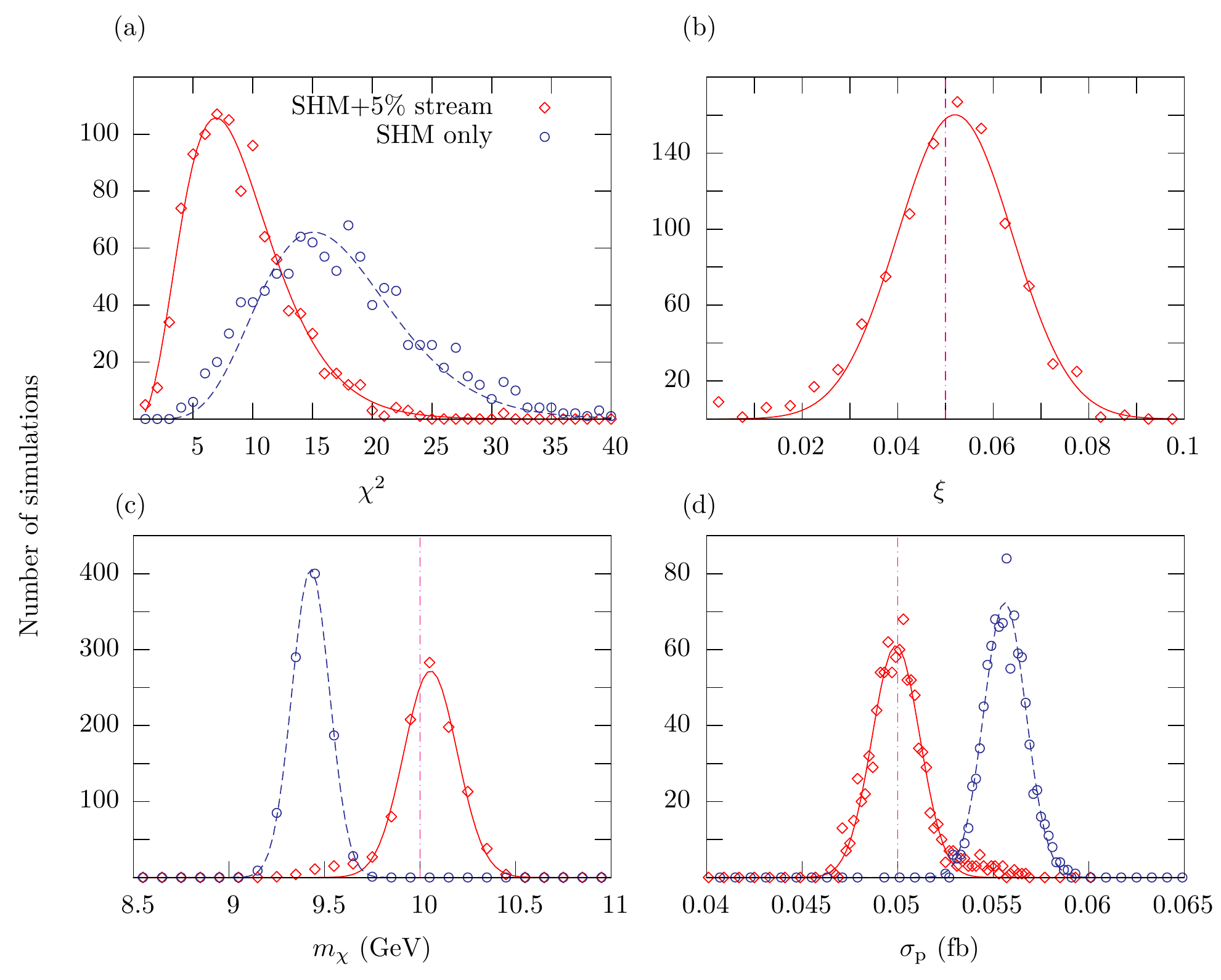}}
\end{center}
\caption{
  Monte Carlo (MC) results and fits for a $\xi$ = 0.05 stream.
  The red(solid) curves show fits to an SHM+stream model, while the
  blue(dashed) curves show fits to an SHM-only model.
  The panels show the distribution of (a) minimized $\chi^2$ and
  best-fit (b) relative stream density $\xi$, (c) dark matter mass
  $m_\chi$, and (d) cross-section $\sigma_{\rm p}$.
  The vertical dashed line indicates the true parameters. (a) is fit by
  a $\chi^2$ distribution, while (b), (c), and (d) are fit by
  Gaussians.\label{fig3} }
\end{figure}

To first examine the impact of a stream on an experimental analysis,
we perform 1000 MC simulations of the CoGeNT results for a given
local stream density $\xi = 0.05$ (see Eq.~\ref{str}). 
We fit each of these simulated results to two types of halo models:
(i) an SHM+stream model with variable $\xi$ and
(ii) an SHM-only model ($\xi = 0$).
Fits are obtained by minimizing the chi-square (using the 10 lowest energy bins in \cite{cogent_new}) over the mass $m_\chi$, cross-section
$\sigma_{\rm p}$, and, for the SHM+stream model, $\xi$.
Fig.~\ref{fig3}(a) shows the minimum chi-square $\chi^2_{\rm min}$ obtained for
fits to the SHM+stream model (red, solid) and the SHM-only model (blue,
dashed).  The SHM+stream model fares significantly better with a
median $\chi^2_{\rm min}$ of 8.2/9 d.o.f.\ compared to a median
$\chi^2_{\rm min}$ of 16.7/10 d.o.f.\ for the SHM-only model. Fig.~\ref{fig3}(b)
shows the best fit values of $\xi$ obtained for the different simulations.
Figs.~\ref{fig3}(c) and (d) show the best fit values of $m_\chi$ and
$\sigma_{\rm p}$, respectively, for the SHM+stream model and the SHM
only model. The SHM+stream model gives best fit values of $m_\chi$ and
$\sigma_{\rm p}$ very close to the true values. The SHM-only model on
the other hand underestimates the mass by $\approx 6\%$ and
overestimates the cross section by $\approx 11\%$. The Sagittarius
stream is clearly visible in (b) and results in erroneous values of
$m_\chi$ and $\sigma_{\rm p}$ if the presence of the stream is
ignored, as in the SHM-only model.

\begin{figure}[!h]
\begin{center}
\scalebox{0.57}{\includegraphics{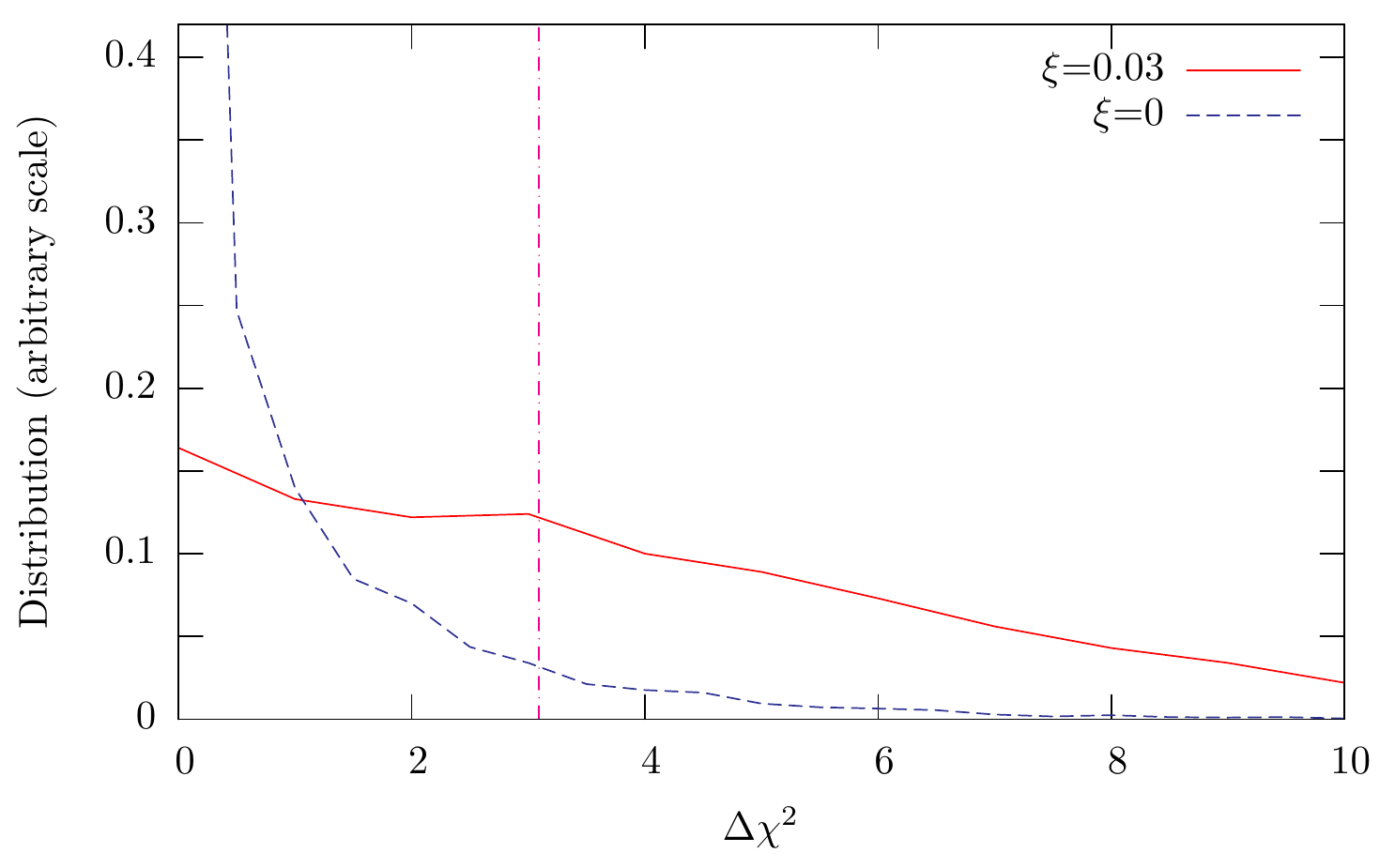}}


\end{center}
\caption{
  The distribution of $\Delta\chi^2$ for an SHM-only halo (dashed blue)
  and an SHM+stream halo with $\xi = 0.03$ (solid red) as determined
  from MC simulations (see the text). The vertical line indicates the median value of $\Delta\chi^2$ = 3.1 for the SHM+stream halo. The median value is exceeded in
  only 5.3\% of the SHM-only simulations.
  \label{fig4} }
\end{figure}

\begin{figure}[!h]
\begin{center}
\scalebox{0.55}{\includegraphics{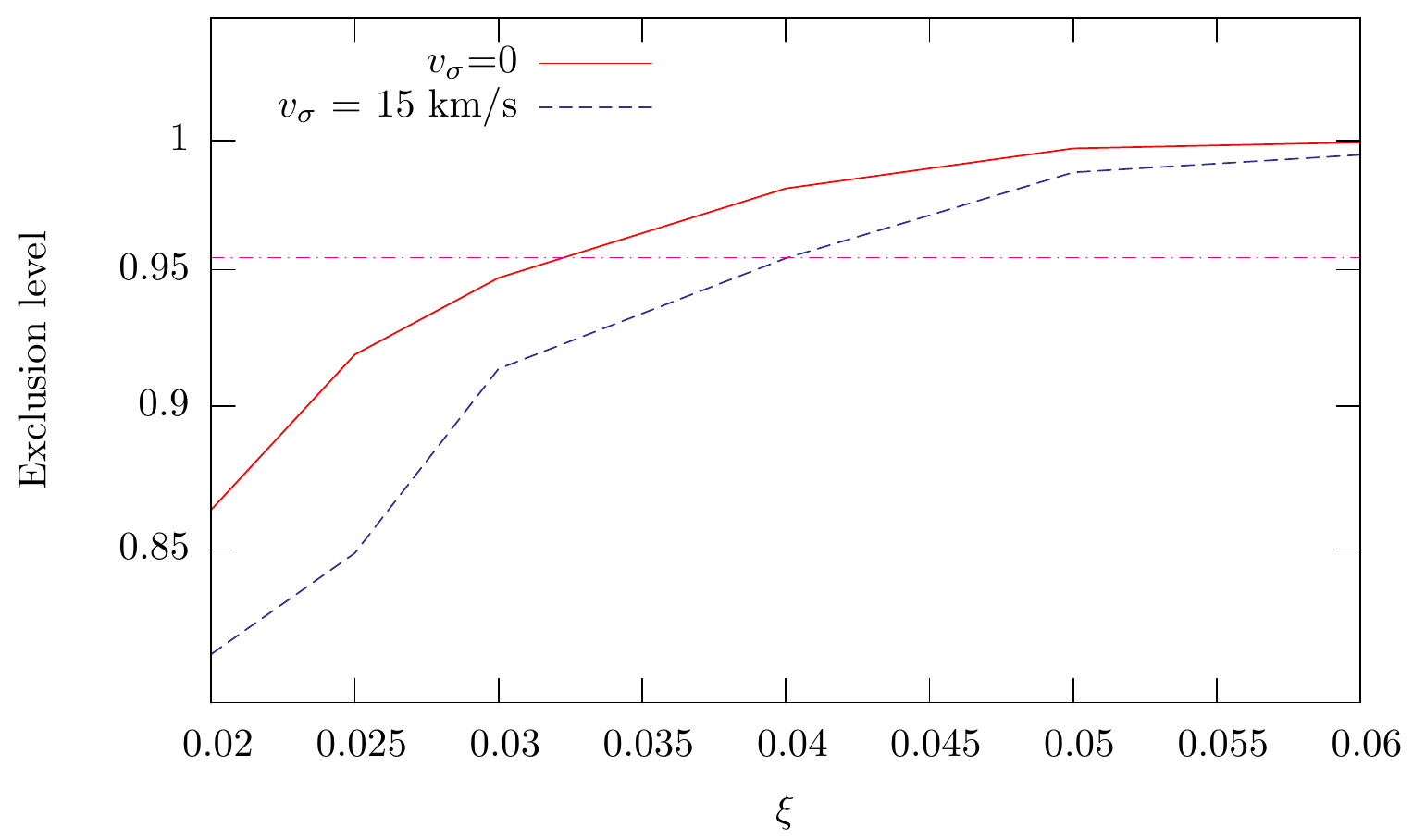}}
\end{center}
\caption{ Confidence level at which the SHM-only model may be excluded
  when the true halo contains a stream of density fraction $\xi$.
  Exclusion levels are shown for streams with velocity dispersions
  $v_\sigma = 0$ (solid red) and 15~km/s (dashed blue).
  \label{fig5} }
\end{figure}

To quantify the ability of CoGeNT to exclude the SHM-only halo model in
favor of a halo also containing the Sagittarius stream, we apply a
likelihood ratio test, which in this case is equivalent to examining
the statistic $\Delta\chi^2 = \chi^2_{\rm min}\textrm{(SHM-only)}
 - \chi^2_{\rm min}\textrm{(SHM+stream)}$, where the two $\chi^2_{\rm min}$
are the minimum chi-square obtained using an SHM-only halo (minimized
over $m_\chi$ and $\sigma_{\rm p}$ with fixed $\xi = 0$) and an
SHM+stream halo (minimized over $m_\chi$, $\sigma_{\rm p}$, and $\xi$),
respectively.
The distribution of this statistic for an SHM-only true halo model,
as determined from 10,000 MC simulations, is shown in Fig.~\ref{fig4} (dashed
blue).  The distribution falls off rapidly for $\Delta\chi^2$ above
$\sim 1$, similar to a $\chi^2$-distribution with 1~d.o.f. \footnote{
  While the likelihood ratio test statistic is expected to follow a
  $\chi^2$-distribution in many cases, we note that the distribution
  here is not actually very well fit by any $\chi^2$-distribution due to
  the physicality condition $0 \le \xi \le 1$, which leads to a pile-up
  at $\xi = 0$ in the fits (42\% of the simulations are best fit by
  $\xi = 0$).}.
The distribution is peaked about small $\Delta\chi^2$ as adding a stream
component to the fit is not expected to significantly improve the
$\chi^2_{\rm min}$ over the SHM-only fit.
Also shown in Fig.~\ref{fig4} is the distribution of $\Delta\chi^2$ assuming the
true halo also contains the Sagittarius stream with $\xi = 0.03$ (solid
red), as determined from 1000 MC simulations.
Including a stream in the fit now allows a substantial improvement in
the $\chi^2_{\rm min}$ over an SHM-only fit, leading to a much broader
$\Delta\chi^2$ distribution.  The median $\Delta\chi^2$ is
3.1 for this halo, whereas only 5.3\% of the simulations yield
$\Delta\chi^2 \ge 3.1$ when the true halo is SHM-only.
In this case, 50\% of the time, the CoGeNT results can be expected to
exclude the SHM-only halo in favor of an SHM+stream halo at the 94.7\%
confidence level (CL).

Fig.~\ref{fig5} shows the CL at which a typical CoGeNT result can exclude the
SHM-only halo as a function of $\xi$, the true stream density.  The
exclusion level is shown for streams with velocity dispersions of
$v_\sigma = 0$ (solid red) and 15~km/s (dashed blue).  The typical
CoGeNT result is defined as the median $\Delta\chi^2$ as determined from
MC simulations.  In other words, there is a 50\% chance that the CoGeNT
results will exclude the SHM-only model at the given CL or better.
The horizontal dashed line represents the $2\sigma$ level. Thus, the
Sagittarius stream is detectable at $>2\sigma$ with a $\sim$ 10~kg-year
exposure with CoGeNT, provided the velocity dispersion associated with
the stream is low, and the stream contributes 3-5\% of the local dark
matter density.

\section{Discussion}
In this paper, we studied the ability of a future CoGeNT data set to
detect the presence of dark matter streams. We performed Monte Carlo
simulations of a halo that consists of both a thermal component, and a
cold stream, and fitted 2 models to the simulations: (i) a halo model
containing the stream and (ii) the SHM-only model. We then performed
simulations of a fully thermal halo (i.e. SHM-only), and fitted the 2
models to the null simulations. We studied the Sagittarius stream as
an example, and showed that for stream densities $\sim 3-5\%$ of the
local dark matter density, the stream is detectable at the $2\sigma$
level with an exposure of 10 kg year. Such an exposure is attainable
by CoGeNT C-4 within $\sim$ 3 years. We set the particle mass = 10
GeV, and assumed knowledge of the stream velocity. Let us now briefly
consider variations in these parameters. \\

\emph{Varying the particle mass:} 
The particle mass $m_\chi$=10 GeV provides an acceptable fit to the
CoGeNT observation, but lies at the high end of the mass range for
$v_0$ = 220 km/s. As the mass is lowered, we lose sensitivity to the
stream (for $v_0$ = 220 km/s) and for $m_\chi < 8$ GeV, the Sagittarius stream becomes almost
completely invisible as recoil events fall entirely
below the energy threshold of 0.47 keVee. This is however, dependent on the assumed values of $v_0$ and $v_{\rm esc}$ \cite{massv0}. We have verified with our simulations that for $m_\chi \geq 8.5$ GeV, we are
able to detect the presence of the stream. Other high velocity streams should be visible for smaller WIMP masses.  \\

\emph{Varying the stream parameters:}
In order to test the importance of our knowledge of the stream
parameters, we perform fits with a random component added to the
stream velocity. The stream speed relative to the sun is chosen at
random to lie between $\pm$50 km/s from the true value, while the two
angles that describe the stream arrival direction are chosen to lie
between $\pm$20 degrees of the true direction. This represents a small
uncertainty in our knowledge of the stream parameters.  We performed
fits to 1000 MC simulations of the SHM+5\% Sagittarius stream, with 25
such random velocities, and obtained $\chi^2_{\rm min}$ values ranging
from 8.4 to 12.1, with a median value of 9.3 with 9 d.o.f. By
comparison, knowledge of the true stream parameters yielded a
$\chi^2_{\rm min}$ of 8.2/9 d.o.f. The SHM-only model resulted in a
substantially worse fit (for $\xi$=0.05), with a $\chi^2_{\rm min}$ of
16.7/10 d.o.f. We thus conclude that an approximate knowledge of the
stream parameters is still useful when analyzing data from
experiments.

We have shown that dark matter streams are potentially detectable by
the future CoGeNT C-4 experiment. Ignoring the presence of streams may
result in erroneous estimates of $m_\chi$ and $\sigma_{\rm p}$. For
sufficiently large exposures, the annual modulation provides
additional information. The annual modulation is an excellent
indicator of streams \cite{str1, str2}, and should reveal the presence of the
stream at a measured energy near the cutoff energy of the
stream. Reconstructing the stream parameters for arbitrary dark matter
streams using a combination of the percentage modulation and the total
number of recoils will enable us to understand the phase space
distribution of dark matter in the solar neighborhood. In previous work 
\cite{str1, str2}, we showed that streams can alter the
phase and structure of the annual modulation. In certain energy bins, the presence of the stream may only be apparent during part of the year, when the stream speed relative to the earth is largest. In these energy bins, the annual modulation due to a $\sim$ few percent stream may result in an annual modulation of a few percent, comparable to the contribution of the entire Maxwellian halo. Observing the variation in the amplitude of the annual modulation in different energy bins provides valuable information regarding the particle mass \cite{form_factor2, wimp_mass1, wimp_mass2, green2, wimp_mass3}. The phase of CoGeNT's annual modulation does not fit the SHM very well and in
our next analysis we will investigate whether  streams lead to a better fit.
We plan to undertake this work at a later stage.

\acknowledgments{We thank Juan Collar and Simon White for helpful discussions.
  A.N.\ thanks the Bruce and Astrid McWilliams Center for Cosmology for
  financial support.
  C.S.\ is grateful for financial support from the Swedish Research
  Council (VR) through the Oskar Klein Centre.
  This research is  supported (KF) in
part by Department of Energy (DOE) grant  DE-FG02-95ER40899 and  by the Michigan Center
for Theoretical Physics.  
KF thanks the Texas Cosmology Center (TCC) where she was a Distinguished Visiting Professor. TCC is supported by the College of Natural Sciences and the Department of Astronomy at the University of Texas at Austin and the McDonald Observatory. KF  also thanks the  Aspen Center for Physics for hospitality during her visit.

  }

\end{document}